\documentclass[fleqn,usenatbib]{mnras}

\usepackage[T1]{fontenc}
\usepackage{ae,aecompl}

\usepackage{graphicx}	
\usepackage{amsmath}	
\usepackage{amssymb}	

\usepackage{gensymb}
\usepackage{array} 
\usepackage{subcaption}	
\newcolumntype{L}[1]{>{\raggedright\let\newline\\\arraybackslash\hspace{0pt}}m{#1}}
\newcolumntype{C}[1]{>{\centering\let\newline\\\arraybackslash\hspace{0pt}}m{#1}}
\newcolumntype{R}[1]{>{\raggedleft\let\newline\\\arraybackslash\hspace{0pt}}m{#1}}
\newcommand{\comment}[1]{}

\title[Constraining activity distribution]{Asteroseismic constraints on active latitudes of solar-type stars: HD\,173701 has active bands at higher latitudes than the Sun}

\author[A. Thomas et al.]{Alexandra E. L. Thomas$^{1,2}$,\thanks{E-mail: axt367@student.bham.ac.uk (AELT)}
William J. Chaplin$^{1,2}$,
Guy R. Davies$^{1,2}$,
\newauthor{Rachel Howe$^{1,2}$,}
\^{A}ngela R. G. Santos$^{3}$,
Yvonne Elsworth$^{1,2}$,
Andrea Miglio$^{1,2}$,
\newauthor{Tiago Campante$^{4,5}$},
Margarida S. Cunha$^{4}$
\\
$^{1}$School of Physics and Astronomy, University of Birmingham, Edgbaston, Birmingham, B15 2TT, UK\\
$^{2}$Stellar Astrophysics Centre, Department of Physics and Astronomy, Aarhus University, Ny Munkegade 120, DK-8000 Aarhus C, Denmark\\
$^{3}$Space Science Institute, 4750 Walnut Street, Suite 205, Boulder, CO 80301, USA\\
$^{4}$Instituto de Astrof\'{\i}sica e Ci\^{e}ncias do Espa\c{c}o, Universidade do Porto, Rua das Estrelas, PT4150-762 Porto, Portugal\\
$^{5}$ Departamento de F\'{\i}sica e Astronomia, Faculdade de Ci\^{e}ncias da Universidade do Porto, Rua do Campo Alegre, s/n, PT4169-007 Porto, Portugal
}

\date{Accepted XXX. Received YYY; in original form ZZZ}

\pubyear{2019}

\begin{document}
\label{firstpage}
\pagerange{\pageref{firstpage}--\pageref{lastpage}}
\maketitle

\begin{abstract}
We present a new method for determining the location of active bands of latitude on solar-type stars, which uses stellar-cycle-induced frequency shifts of detectable solar-like oscillations. When near-surface activity is distributed in a non-homogeneous manner, oscillation modes of different angular degree and azimuthal order will have their frequencies shifted by different amounts. We use this simple concept, coupled to a model for the spatial distribution of the near-surface activity, to develop two methods that use the frequency shifts to infer minimum and maximum latitudes for the active bands. Our methods respond to the range in latitude over which there is significant magnetic flux present, over and above weak basal ephemeral flux levels. We verify that we are able to draw accurate inferences in the solar case, using Sun-as-a-star helioseismic data and artificial data. We then apply our methods to \emph{Kepler} data on the solar analogue HD\,173701, and find that its active bands straddle a much wider range in latitude than do the bands on the Sun.
\end{abstract}




\section{Introduction}
\label{sect: intro}
Spatially resolved observations of the Sun reveal that regions of strong surface magnetic activity are not distributed uniformly over the solar surface, but instead lie in so-called active bands of latitude either side of the equator \citep{1858MNRAS..19....1C}. These bands migrate during the solar cycle and this can be observed by mapping surface magnetic fields over time \citep[e.g.][]{1981SoPh...74..131H} or by resolving the solar surface and studying the positions of sunspots giving rise to the classic butterfly diagram \citep[e.g.][]{1904MNRAS..64..747M}. The distribution of sunspots is a manifestation of the action of the Sun's internal dynamo, which converts poloidal into toroidal field during the rising phase of the activity cycle. Concentrated field lines form in bands parallel to the equator and, as the cycle progresses, move towards lower latitudes.  It is reasonable to assume that other solar-type stars are likely to show active bands of latitude. 

Some techniques exist to map the topology of magnetic activity on stellar surfaces. One method uses spot-induced modulations in stellar lightcurves to measure surface differential rotation and, if the inclination angle of the star is known, to constrain the latitudes of the spots \citep[e.g.][]{1979ApJ...227..907E,2003A&A...403.1135L,2007A&A...464..741L,2003ApJ...585L.147S,2017A&A...599A...1S}. Tests on artificial data \citep{2015MNRAS.450.3211A} suggest that results from such analyses may need to be treated with caution. Other methods include the Zeeman Doppler imaging technique \citep[][and references therein]{2014IAUS..302..180F} and spectro-polarimetric observations \citep{2002A&A...381..736P,2014A&A...565A..83K}. These have been applied to Ap and Bp stars \citep[e.g.][]{2015A&A...573A.123R,2011ApJ...726...24K} as well as to Sun-like exoplanet hosts \cite[e.g.][]{2013MNRAS.435.1451F,2015A&A...582A..38A}.

\citet{2007MNRAS.377...17C} and \citet{2014aste.book....1C} suggested that it should be possible to use stellar-cycle-induced frequency shifts of solar-like oscillations in Sun-like stars to place constraints on the latitudinal dependence of the near-surface magnetic activity. That frequencies of solar acoustic (p) modes vary with the solar cycle is now a well established result \citep[e.g.][]{1990Natur.345..779L,1990Natur.345..322E,1998A&A...329.1119J,2002ApJ...580.1172H,2004MNRAS.352.1102C,2007ApJ...659.1749C}.  Temporal changes of p-mode oscillation frequencies have now been observed in several Sun-like stars \citep[e.g.][]{2010Sci...329.1032G,2016A&A...589A.118S,2016A&A...589A.103R,2017A&A...598A..77K,2018A&A...611A..84S,2018ApJS..237...17S} thanks to the availability of long, high-cadence, high-quality lightcurves from the French-led CoRoT satellite, and in particular the NASA \emph{Kepler} Mission. This includes the solar analogue HD\,173701 (HIP\,91949, KIC\,8006161), which is the focus of this paper. With near-solar radius, mass and age, it exhibits solar-like differential rotation and a $7.4$-year activity cycle, making it a particularly interesting object for comparison with the Sun \citep{2018ApJ...852...46K}. Its p-mode frequency shifts have been found to be larger than those seen in the Sun \citep[e.g.][]{2017A&A...598A..77K}, which \citeauthor{2018ApJ...852...46K} attributed to a higher metallicity and a deeper outer convection zone. 

The layout of the rest of the paper is as follows. We begin in Section~\ref{sect: general theory} by explaining the basic principles of using frequency shifts of solar-like oscillations to constrain the location of the active latitudes on a star. As we shall see, the technique has the potential to provide particularly useful constraints for dynamo modellers, since it responds to the latitude range over which significant magnetic flux is present over and above basal, ephemeral levels. 

In order to validate our approach we used Sun-as-a-star helioseismic data and artificial asteroseismic data, before applying the methodology to over 3\,years of \emph{Kepler} data on HD\,173701. The data we used are described in Section~\ref{sect: data}. We then present in detail in Section~\ref{sect: methodology} the two complementary ways in which we used the frequency shift data to infer the active latitudes. The results from verification using solar and artificial data are presented in Section \ref{subsect: Sun results}. We then apply the same methods to the frequencies of HD\,173701 (Section~\ref{subsect: doris results}), which was observed during the nominal \textit{Kepler} Mission. We conclude our paper with a summary of the key findings in Section~\ref{sect: conclusions}.

\section{Basic principles of the method}
\label{sect: general theory}
Sun-like stars show a rich spectrum of overtones of solar-like oscillations, acoustic pulsations that are stochastically excited and intrinsically damped by near-surface convection. Since gradients of pressure act to restore the modes, they are often colloquially referred to as p modes. The spatial properties of the modes may be described by spherical harmonics of angular degree $l$ and azimuthal order $m$. Any spherical asymmetry, such as rotation or magnetic fields, can lift the degeneracy in $l$, so that each non-radial mode is split into $2l+1$ azimuthal components, where $-l \le m \le l$. Overtones of each $l$ are labelled by a radial order, $n$.

The frequencies of the modes shift in the presence of magnetic activity, either due to the direct effects of the Lorentz force or indirectly from magnetic-field-induced changes in the sound speed within the mode cavities. With a non-homogeneous distribution of magnetic activity, modes of different angular degree and azimuthal order will experience different frequency shifts \citep{2000MNRAS.313..411M}. The various combinations of $l$ and $m$ have different spatial distributions on a sphere and therefore the responses, in amplitude and phase, of these components depend on the magnetic field strength in the regions where the acoustic waves forming them propagate. For this work we assume that magnetic and spin axes are aligned.

Under the assumption that the source driving the sound-speed perturbations, and therefore the shifts in frequency, is located close to the stellar surface, those shifts may be described by:
 \begin{equation}
 \delta\nu_{lm} \propto \: \Big(l+\frac{1}{2}\Big) \: \frac{(l-|m|)!}{(l+|m|)!}  \int\limits_{\theta_{\rm min}}^{\theta_{\rm max}}|P^{|m|}_{l}\:(\cos\theta)|^2 \: B(\theta) \: \sin\theta \: d\theta,
 \label{eqn: freq shift}
 \end{equation}
where the co-latitude $\theta=(\frac{\pi}{2}-\lambda)$ and $\lambda$ the latitude, $B(\theta)$ is the magnetic field strength, and $P^{|m|}_{l}(\cos\theta)$ are the associated Legendre polynomials \citep{2000MNRAS.313..411M}. Note how the magnitude of the shift depends on ($l$,$|m|$). The relative sizes of the frequency shifts therefore give information on the spatial distribution of the near-surface activity.

From highly spatially resolved observations of the Sun over time it is possible to measure the frequency shifts of modes of individual ($l$,$m$) up to degrees that provide high-fidelity constraints on the spatial properties of the near-surface activity. For example, \cite{2002ApJ...580.1172H} used frequency shifts of modes covering the range $l=0$ to 150 observed by the Global Oscillations Network Group (GONG) -- with effectively a separate frequency available for each azimuthal order $m$ -- to produce a map of levels of shift distributed over latitude and time. This map looks very similar to butterfly diagrams that show spatial and temporal changes in levels of magnetic field.

For unresolved observations of stars, geometric cancellation means we are unable to obtain data over such a wide range in $l$. Using intensity measurements from \textit{Kepler} it is possible to detect only overtones of $l=0$, 1 and 2, and in a few cases $l=3$, and so we must accept that any latitudinal localisation from asteroseismology will be less refined than is possible in the resolved-Sun case. Nevertheless, shifts of varying size have been observed for different low-degree solar p modes by \cite{2004MNRAS.352.1102C}, \cite{2004ApJ...610L..65J}, \cite{2007ApJ...659.1749C} and \cite{2015A&A...578A.137S} that are entirely consistent with the non-spherically-symmetric distribution of near-surface activity on the Sun. In what follows we make use of data on $l=0$ and $l=1$ modes, since $l=2$ frequency shifts inferred from \emph{Kepler} data are noticeably noisier (owing to their lower visibility in unresolved observations compared to the radial and dipole modes).

For simplicity, and given the above constraints, we model the magnetic field distribution as a top-hat function assuming bands of activity wrapped around a star parallel to its equator, i.e.,
\begin{equation}
B(\theta=\frac{\pi}{2}-\lambda)=\begin{cases}
b, & \text{if } \lambda_{\rm min}\leq\lambda\leq\lambda_{\rm max},\\
0, & \text{otherwise},
\end{cases}
\label{eqn: B-field}
\end{equation}
where $\lambda_{\rm min}$ and $\lambda_{\rm max}$ are the minimum and maximum latitudes that define the band in each hemisphere -- the southern hemisphere is assumed to be a reflection of the northern hemisphere since globally coherent modes provide an averaged measure of the properties above and below the stellar equator -- and $b$ is a constant which is independent of latitude. Using  butterfly diagrams built from observations of the Sun we find evidence of solar activity distributions lying along such bands. We therefore use Equation~\ref{eqn: B-field} as an approximation of activity distribution which can be applied to other Sun-like stars. 

In what follows we attempt to constrain distributions that are integrated over long epochs (entire rising or falling phases, or over an entire cycle). We therefore ignore any cyclic variations where the latitudinal dependence of activity may change across the period of time observed (within the resolution of observation) and instead seek a epoch-averaged distribution. Consequently, $\lambda_{\rm min}$ and $\lambda_{\rm max}$ are time-independent; the precision in the measurements of the frequency shifts does not allow us to constrain more complex models that include time variation of the latitudes. Although we acknowledge that this assumption is not entirely true since bands of activity migrate during the solar cycle, a justification is discussed in a later section. However, since $b$ is representative of the time-varying field strength it is of course time-dependent.  

By measuring the frequency shifts of modes of different ($l$,$m$) over time we may place constraints on $\lambda_{\rm min}$ and $\lambda_{\rm max}$. The frequency shifts needed for the analysis are calculated from estimates of mode frequencies over different epochs. We need observations of a star that span at least a significant fraction of the ascending or descending phase of its activity cycle. The full lightcurve is divided temporally into shorter lightcurves, and frequencies extracted from each to enable time-dependent frequency shifts to be calculated. 

The extraction of individual frequencies proceeds by fitting multi-parameter models to the resonant mode peaks in the frequency-power spectrum of a lightcurve, a process colloquially referred to as peak-bagging \cite[e.g.][]{1990ApJ...364..699A,2011A&A...527A..56H,2015MNRAS.446.2959D,2017ApJ...835..172L}. Because the modes are damped, they manifest as Lorentzian-like peaks in the spectrum. The ability to resolve the individual $m$ components of a given ($n$,$l$) multiplet depends on how the widths of peaks compare to the splittings between them, which components are detectable, and the frequency resolution of the data. The visibility of modes in the power spectrum (assuming photometric observations) is well described by  
\begin{equation}
\varepsilon_{lm}(i) = \frac{\left(l-|m|\right)!}{\left(l+|m|\right)!} \left(P_{l}^{|m|}(\cos i)\right)^2,
\label{eqn: inc}
\end{equation}
where $\varepsilon_{lm}(i)$ is the mode visibility and $P_{l}^{|m|}(\cos i)$ are Legendre Polynomials \citep{2003ApJ...589.1009G}. Depending on the angle of inclination $i$ offered by the star, the above implies that due to geometric cancellation when averaging over the visible stellar disc, the non-spherically symmetric perturbations for some of the modes may have zero visibility. 

Long, multi-year lightcurves give the exquisite frequency resolution needed to resolve individual $m$ components of the non-radial modes. Equation~\ref{eqn: inc} implies that if one can then measure the relative power in the detectable $m$, it is possible to constrain the angle of inclination of the star (for a detailed description see \citet{2015MNRAS.446.2959D}). However, multi-year lightcurves are too long to track stellar cycle changes and their frequency shifts. Instead we must divide the data into subsets of a few month's duration. The reduced frequency resolution then makes it extremely hard to disentangle the frequencies of the individual $m$ components of the non-radial modes. Rather than having the luxury of working with shifts for individual $m$ components, this means that in practice we must use shifts for each non-radial ($n$,$l$) multiplet that are \emph{weighted averages} of the shifts shown by the constituent $m$. The respective weighting depends on the relative visibilities of the different components, as described by Equation~\ref{eqn: inc}.

The usual practice is to fit a single frequency to each non-radial multiplet -- in effect the weighted centroid of the contributing components (see Figure \ref{fig: model multiplet}) -- and a single frequency splitting to describe the separations between adjacent $m$ components. Since the visibility of each $m$ component is dependent on the angle of inclination of the star it is necessary to understand the relative contribution of each of the components to the centroid of the multiplet at different angles, $i$. For dipole ($l=1$) modes, this balance may be described by
\begin{equation}
\nu_{nl} = \alpha \nu_{nl0} + (1-\alpha) \nu_{nl|1|},
\label{eqn: nufit quick}
\end{equation}

where $\nu_{nl}$ is the central frequency of the multiplet, $\nu_{nl0}$ is the frequency of its constituent $m=0$ component, and $\nu_{nl|1|}$ is the average frequency of the $m=-1$ and 1 components. The coefficient $\alpha$ describes the relative contributions of each azimuthal component which depend on the underlying mode parameters and the inclination angle of the star. These were calculated as detailed in Appendix \ref{adx: alpha_i_relation}. 

In summary, provided we have a good estimate from the full dataset of the angle of inclination of the star, we can work with $m$-averaged frequency shifts since we know the relative contribution of the different $m$ to these estimates.

\section{The data}
\label{sect: data}

We used just over 3\,years of short-cadence \textit{Kepler} data on HD\,173701, covering observing quarters Q5.1 through Q17.2 inclusive from the nominal mission. This period spans most of a rising phase of the 7.4-year stellar activity cycle shown by the star \citep{2018ApJS..237...17S}. Lightcurves were corrected using the KASOC filter \citep{2014MNRAS.445.2698H} and prepared as described in \citet{2017ApJ...835..172L}. Frequencies were estimated for 90-day subsets of the full lightcurve (each overlapping by 45 days), as described in \cite{2018ApJS..237...17S}. For reference later in the paper, the inclination angle of HD\,173701 was estimated using the full lightcurve to be $i = 37.2^{+4.0}_{-4.0}$\,degree, using the approach described in \citet{2015MNRAS.446.2959D}; this is consistent with the value quoted in \citet{2018ApJ...852...46K}.

\begin{figure}
	\centering
	\includegraphics[width=0.48\textwidth]{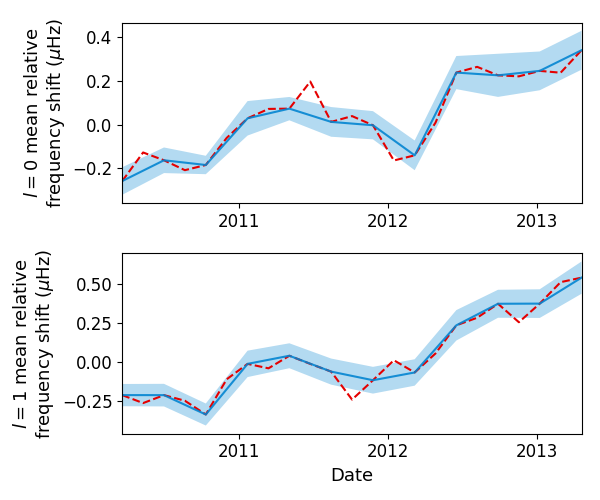}
	\caption{Mean frequency shifts from the 90-day data on HD\,173701, averaged over the six strongest orders at $l=0$ and $l=1$, respectively. All available (and overlapping) data are shown in dashed red and overlaid in solid blue are the independent data sets with no overlap.}
	\label{fig: doris cycle}
\end{figure}

Figure~\ref{fig: doris cycle} shows frequency shifts from these 90-day data, averaged over six orders of, respectively, $l=0$ modes (top panel) and $l=1$ modes (bottom panel), centered on the order showing the highest radial-mode amplitudes. The reasons for choosing these six orders are discussed in Section~\ref{sect: methodology} below. The selected modes span the frequency range $\approx 3100$ to $3900\,\rm \mu Hz$. Results on the overlapping sets are shown in red, whilst the dark blue line (and its lighter 1\,$\sigma$ uncertainty envelope) shows results on the independent data used in the analyses.

\begin{figure}
	\centering
	\includegraphics[width=0.48\textwidth]{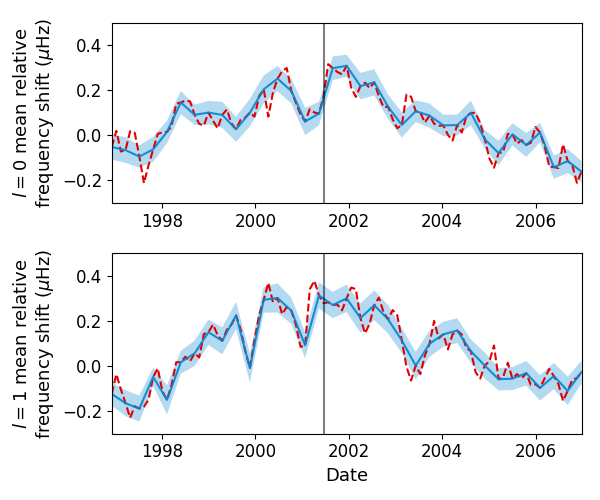}
	\caption{Mean frequency shifts from the 108-day BiSON data on the Sun, averaged over the six strongest orders at $l=0$ and $l=1$, respectively. Data for solar cycle 23 are shown in dashed red and overlaid in solid blue are the independent data sets used for the analyses. The vertical lines indicate the separate rising and falling phases used.}
	\label{fig: whole bison cycle}
\end{figure}

Our methods (described in Section \ref{sect: methodology}) were validated by application to low-$l$ solar p-mode frequencies measured by the Birmingham Solar-Oscillations Network (BiSON) \citep{1996SoPh..168....1C,2016SoPh..291....1H}. BiSON is comprised of six ground based telescopes that observe the Sun as if it were a star in Doppler velocity. Mode frequencies were extracted from 108-day timeseries covering solar cycle 23 using the peak-bagging approach outlined in \citet{2015MNRAS.454.4120H}. This length of timeseries covers four solar rotations and so any signal arising from sun-spot modulation is smoothed over. The timeseries start at 36-day intervals, giving each timeseries a 72-day overlap with adjoining series. Our analysis to infer the active latitudes  was carried out on a full solar cycle as well as on individual rising and falling sections. The chosen epochs are shown in Figure \ref{fig: whole bison cycle}, which displays weighted mean frequency shifts averaged over the six orders at $l=0$ and $l=1$, respectively, centred on the frequency of highest amplitude (which for the Sun spans the frequency range $\approx 2750$ to $3500\,\rm \mu Hz$).

For additional method validation, low-$l$ solar data from the VIRGO (Variability of solar IRradiance and Gravity Oscillations) SPM (sun photometers) on board the Solar and Heliospheric Observatory (SoHO) \citep{1995SoPh..162..101F,1997SoPh..170....1F} were also analysed. Frequencies were estimated from peak-bagging applied to the power spectra of 180-day lightcurves, each overlapping in time by 90 days \citep{2018ApJS..237...17S}. Timeseries of 180 days were used since frequencies from 90-day datasets were notably noisier. The VIRGO data cover approximately 12 years, and hence we were able to analyse one entire 11-year cycle in addition to the constituent rising and falling phases.

\section{Methodology}
\label{sect: methodology}

We used two methods to constrain the active latitudes, both using estimates of individual mode frequencies from different epochs as inputs. The first method works directly with the ratios of shifts shown by non-radial and radial modes.

\subsection{Method 1: Ratio of frequency shifts}
\label{sect: method1}
From Equation \ref{eqn: freq shift} we know that observed frequencies will experience a shift due to the presence of a magnetic field. This can be described as a shift with respect to a field-free epoch such that at some epoch, $t_j$, the measured frequency of a mode having radial order $n$ and degree $l$ is:
 \begin{equation}
 \nu_{nl}(t_j)=\nu_{nl}(t_0)+\delta\nu_{nl}(t_j),
 \label{eqn: freq}
 \end{equation}
where frequencies for non-radial modes are the aforementioned weighted averages over the constituent $m$ components (Equation~\ref{eqn: nufit quick}), and $\nu_{nl}(t_0)$ are reference, field-free frequencies. Once extracted, the frequency shifts $\delta\nu_{nl}(t_j)$, can be used to constrain the latitudinal dependence.  A credible and obvious concern with the above is how one would in practice go about estimating a field-free frequency, i.e., even at solar minimum, significant amounts of widely distributed magnetic flux are still present in the near-surface layers, implying the mode frequencies will be offset from those expected for a notional field-free Sun. Provided the surface field architecture (i.e., the latitudinal dependence) does not change significantly over time, or instead the quality of the observations precludes us from detecting such changes (so in effect we see the same distribution, irrespective of epoch), the above concern is irrelevant and we can recover the sought-for latitudinal dependence. There is no need to measure shifts with respect to a field-free epoch and instead we may use any ratio of frequency differences, as we now go on to show.

By extracting the frequency shifts, the ratios at a given epoch of the shifts shown by modes of different ($l,m$) should provide constraints on the near-surface latitudinal dependence of activity. The assumption of a constant temporal latitudinal dependence means that the ratio of contemporaneous shifts at two different degrees is always the same, regardless of the epoch. Taking the example of the ratio between $l=1$ and $l=0$ modes of the same radial order at a particular epoch, $t_j$, we have 
\begin{equation}
\eta = \frac{\delta\nu_{n1}(t_j)}{\delta\nu_{n0}(t_j)} = \frac{\delta\nu_{n1}(t_{j+1})}{\delta\nu_{n0}(t_{j+1})} = \frac{\delta\nu_{n1}(t_{j+j'})}{\delta\nu_{n0}(t_{j+j'})},
\label{eqn: eta start}
\end{equation}
where $t_{j+j'}$ is some epoch falling after $t_j$, and $\eta$ is a constant describing the ratio since it is $n$-independent. 

Now, the difference in frequency of a mode at two different epochs is:
\begin{equation}
\nu_{nl}(t_{j+1}) - \nu_{nl}(t_{j}) = \delta\nu_{nl}(t_{j+1}) - \delta\nu_{nl}(t_{j}),
\end{equation}
The ratio of differences in the measured frequencies, and also in the corresponding shifts, is:
\begin{equation}
\Phi_{10}(n,t_j) = \frac{\nu_{n1}(t_{j+1}) - \nu_{n1}(t_{j})}{\nu_{n0}(t_{j+1}) - \nu_{n0}(t_{j})} = \frac{\delta\nu_{n1}(t_{j+1}) - \delta\nu_{n1}(t_{j})}{\delta\nu_{n0}(t_{j+1}) - \delta\nu_{n0}(t_{j})}.
\label{eqn: ratio of shifts phi}
\end{equation}
We can rearrange Equation \ref{eqn: ratio of shifts phi} as follows to obtain $\eta$:
\begin{equation}
\begin{aligned}
\Phi_{10}(n,t_j)\delta\nu_{n0}(t_{j+1}) - \Phi_{10}(n,t_j)\delta\nu_{n0}(t_{j}) \\
= \delta\nu_{n1}(t_{j+1}) - \delta\nu_{n1}(t_{j}), \\
= \eta(\delta\nu_{n0}(t_{j+1}) - \delta\nu_{n0}(t_{j})). \\
\Rightarrow \eta = \frac{\Phi_{10}(n,t_j)(\delta\nu_{n0}(t_{j+1}) - \delta\nu_{n0}(t_{j}))}{\delta\nu_{n0}(t_{j+1}) - \delta\nu_{n0}(t_{j})} \equiv \Phi_{10}(n,t_j).
\end{aligned}
\label{eqn: eta final}
\end{equation}
This shows that by taking any ratio of frequency differences we need not know the shift with respects to a field-free frequency.

Another advantage of taking ratios of modes of a particular $n$ at any epoch can be seen by re-writing the basic equation for $\eta$ with the contributing shifts of the different $l=1$ mode components $\delta\nu_{nlm}$ included explicitly, i.e.,
\begin{equation}
\eta = \frac{\delta\nu_{n1}(t_j)}{\delta\nu_{n0}(t_j)} = \frac{\alpha\delta\nu_{n10}(t_j) + (1-\alpha)\nu_{nl|1|}}{\delta\nu_{n00}(t_j)},
\label{eqn: method1 theoretical eta}
\end{equation}
where $\alpha$ has the same meaning as in Equation \ref{eqn: nufit quick}. Using Equation \ref{eqn: freq shift} and by assuming a given model for the magnetic field (Equation \ref{eqn: B-field}) we again see that the field strength cancels leaving the ratio to depend only on the geometry of the near-surface activity.  Hence the ratios can be used to place constraints on $\lambda_{\rm min}$ and $\lambda_{\rm max}$.

Finally, we also assume that the shifts do not change with radial order (frequency). Solar frequency shifts do increase in size at higher orders. However, provided we limit ourselves to using only a few orders, the typical precision in the data means the above is a reasonable approximation. The frequency dependence of the shifts for HD\,173701 has been shown to be comparable to the Sun \citep{2018A&A...611A..84S}. We tested the impact of including in our model a power law relation in frequency to describe the shifts, as given by \cite{2001MNRAS.324..910C}. We found that adopting a range of reasonable values for the exponent made little impact to the resulting latitude estimation.

Here, we use data on six radial orders for our analysis, straddling the order showing the largest measured mode amplitudes.

\subsubsection{Calculating observed $\eta$}
\label{sect: observed_eta}

Frequency shifts were calculated by taking differences between observed frequencies separated by $k$ epochs, i.e.,
\begin{equation}
\delta\nu_{nl}(t_j) = \nu_{nl}(t_{j+k}) - \nu_{nl}(t_j).
\label{eqn: method 1 shift}
\end{equation}

To optimize the signal-to-noise ratio in the computed shifts, we chose $k=9$ for HD\,173701, which is equivalent to epochs separated by $9\times45$ days. We note that three years of data on HD\,173701, from the total 7.4 year activity cycle \citep{2018ApJ...852...46K} is equivalent to 4.4\,years from the 11 year solar cycle of BiSON data. For BiSON's 108-day data this equated to $k=17$ epochs taken every 36 days. For the 180-day observations of VIRGO, data sets overlapped by 90 days, therefore the corresponding number of epochs was $k=4$.

The ratio between $l=1$ and $l=0$ shifts of radial order $n$, calculated over an epoch difference $j$ is given by
\begin{equation}
\eta_{\rm obs,nj} = \frac{\delta\nu_{n1,j}}{\delta\nu_{n0,j}},
\label{eqn: eta_obs}
\end{equation}
and is evaluated between each pair of modes of the same radial order for all overtones and over all pairs of epochs.  Since ratios of small numbers can be noisy with the available computational precision we perform calculations in logarithmic space, i.e.
\begin{equation}
\ln(\eta_{\rm obs,nj}) = \ln(\delta\nu_{n1,j}) - \ln(\delta\nu_{n0,j}).
\end{equation}
Uncertainties on this logarithmic ratio, $\sigma \ln(\eta_{\rm obs,nj})$, were propagated from observational uncertainties on the individual frequencies used to construct each shift.

To find a time- and order- averaged observed ratio $\left< \eta_{\rm obs}\right>$, we compute the weighted average over all ratios and epochs. The first sum is over all ratios at a given epoch, i.e.,
\begin{equation}
\ln(\eta_{\rm obs,j}) = \left(\sum\limits_{n=1}^{N_{\rm ord}} \frac{\ln(\eta_{\rm obs,nj})}{\sigma \ln(\eta_{\rm obs,nj})^2} \right) / \left( \sum\limits_{n=1}^{N_{\rm ord}}\frac{1}{\sigma \ln(\eta_{\rm obs,nj})^2} \right),
\end{equation}
with an uncertainty given by
\begin{equation}
\sigma \ln(\eta_{\rm obs,j}) = \left(\sum\limits_{n=1}^{N_{\rm ord}}\frac{1}{\sigma \ln(\eta_{\rm obs,nj})^2} \right)^{-1/2},
\end{equation}
where $N_{\rm ord}=6$ is the total number of orders used. The second sum is then made over all epochs, $N_{\rm e}$, i.e.,
\begin{equation}
\ln(\left< \eta_{\rm obs} \right>) = \left( \sum\limits_{j=1}^{N_{\rm e}} \frac{\ln(\eta_{\rm obs,j})}{\sigma \ln(\eta_{\rm obs,j})^2} \right) / \left( \sum\limits_{j=1}^{N_{\rm e}} \frac{1}{\sigma \ln(\eta_{\rm obs,j})^2} \right),
\end{equation}
with an uncertainty given by
\begin{equation}
\sigma \ln(\left< \eta_{\rm obs} \right>) = \left(\sum\limits_{j=1}^{N_{\rm e}}\frac{1}{\sigma \ln(\eta_{\rm obs,j})^2} \right)^{-1/2}.
\end{equation}
The resulting $\left< \eta_{\rm obs} \right>$ depends on the latitudinal profile of the near-surface magnetic activity.

\subsubsection{Constraining the minimum and maximum latitudes}
\label{sect: constrain_lat}

In order to interpret the observed $\left< \eta_{\rm obs} \right>$ we need to identify the model that best matches the observations. From our description of the frequency shifts given by Equation \ref{eqn: freq shift}, it is clear that $\left< \eta_{\rm obs} \right>$ will depend on the magnetic field geometry and hence on $\lambda_{\rm min}$ and $\lambda_{\rm max}$, but not on the field strength since it cancels. A grid of models was constructed using combinations of the parameters $\lambda_{\rm min}$ and $\lambda_{\rm max}$ each in the range zero to 90\,degrees in increments of 0.5\,degrees. Modelled ratios, $\eta_{\rm mod}$, were evaluated using Equation \ref{eqn: method1 theoretical eta} where $\alpha$  was deduced from the known inclination angle and shifts were calculated using Equation \ref{eqn: freq shift} and $\lambda_{\rm min}$ and $\lambda_{\rm max}$. The best-fitting model ratio $\eta_{\rm mod}$ to the observed value $\left< \eta_{\rm obs} \right>$ was identified from the grid of modelled ratios as that which gave the lowest chi-squared statistic.

To build a probability distribution of inferred latitudes we sampled from a distribution that is representative of the observations and iterated multiple times. The measurement of $\left< \eta_{\rm obs} \right>$ and its associated uncertainty, $\left< \sigma\eta_{\rm obs} \right>$, were used to randomly sample from a Gaussian distribution of $\mathcal{N}(\left< \eta_{\rm obs} \right>,\sigma\left< \eta_{\rm obs} \right>)$. From this we produced sample ratios $\eta'_{\rm obs}$ given by
\begin{equation}
\eta'_{\rm obs} = \left< \eta_{\rm obs} \right> + n_i \: \sigma\left< \eta_{\rm obs} \right>,
\label{eqn: eat obs sample}
\end{equation}
where $n_i$ is a random number from a Gaussian distribution with $\mathcal{N}(0,1)$. To incorporate the uncertainty on inclination angle, $\sigma i$, we again sampled from a representative distribution, $\mathcal{N}(i,\sigma i)$, to produce samples $i'$ which were used to calculate each new grid of ratios, $\eta'_{\rm mod}$, i.e.
\begin{equation}
\eta_{\rm mod}' = \frac{\alpha' \delta\nu_{n10} + (1 - \alpha')\delta\nu_{n1|1|}}{\delta\nu_{n00}},
\end{equation}
where $\alpha'$ is calculated using $i'$ (see Appendix \ref{adx: alpha_i_relation}).

For each iteration, where a new sample of $\eta'_{\rm obs}$ was taken and a new grid, $\eta_{\rm mod}'$, was calculated, we minimised $\chi^2$ to find the best fitting $\lambda_{\rm min}$ and $\lambda_{\rm max}$, where
\begin{equation}
\chi^2 = \frac{(\eta'_{\rm obs} - \eta_{\rm mod}')^2}{\eta_{\rm mod}'}.
\end{equation}
Posterior distributions in $\lambda_{\rm min}$ and $\lambda_{\rm max}$ were made after performing 15,000 iterations.

\subsection{Method 2: Forward modelling approach}
\label{sect: method2}

To ensure the robustness of our results we also applied a second method comprising a forward modelling approach.

\subsubsection{The model}
\label{sect: method2, the model}
We directly modelled the frequencies of $l=0$ and $l=1$ modes observed over different epochs. The field-free frequency was therefore included in the model but handled as a nuisance parameter, as described below. We assumed that there was no frequency dependence in the shifts, as discussed in Section \ref{sect: data}.

The frequency shifts in the model are defined by
\begin{equation}
\delta\nu_{lm}(t_j) = \beta(t_j) \, \left(l + \frac{1}{2}\right) \, K_{lm},
\end{equation}
where the $\beta(t_j)$ capture the time dependence of the shifts (see below), and
\begin{equation}
K_{lm} = \frac{(l-|m|)!}{(l+|m|)!} \, \int\limits_{\theta_{\rm min}}^{\theta_{\rm max}} |P^{|m|}_l \,(\cos\theta)|^2 \, \sin\theta d\theta.
\end{equation}
As a reminder, $\theta$ is the co-latitude given by $\theta = \left(\frac{\pi}{2} - \lambda\right)$.

As discussed in Section \ref{sect: general theory}, we extract what are in effect centroid frequencies for each non-radial mode rather than frequencies for the individual $m$ components. The above equations, when combined with Equation \ref{eqn: freq}, then become
\begin{equation}
\nu_{nl}(t_j) = \nu_{nl}(t_0) + \beta(t_j) \, \big(l + \frac{1}{2}\big) \, K_{l},
\end{equation}
and
\begin{equation}
K_l = \alpha K_{l0} + (1-\alpha)K_{l|1|},
\end{equation}
describes the contributions of the different $m$-components to the centroid frequency (using $\alpha$ described in Appendix \ref{adx: alpha_i_relation}).

We re-parametrise so that $\beta(t_j)=\beta_j$ is used to describe the shifts of the $l=0$ and $l=1$ modes from their field-free frequencies and $I(\theta_{\rm min}, \theta_{\rm max}) = K_1/K_0$ acts as a correction factor to incorporate the latitudinal dependence of magnetic field which will only affect non-radial modes. We define $\theta_{\rm min}=\left(\frac{\pi}{2}-\lambda_{\rm min}\right)$ and $\theta_{\rm max}=\left(\frac{\pi}{2}-\lambda_{\rm max}\right)$. The result is that we model radial and dipole frequencies separately, but with shared parameters, as described by
\begin{equation}
\begin{aligned}
\nu_{n0,j} &= \nu_{n0,0} + \frac{1}{2}\beta_{j},\\
\nu_{n1,j} &= \nu_{n1,0} + \frac{3}{2}\beta_{j} I(\theta_{\rm min},\theta_{\rm max}),
\end{aligned}
\label{eqn: method 2 dipole shift}
\end{equation}
and
\begin{equation}
	I(\theta_{\rm min},\theta_{\rm max}) = \dfrac{ \alpha A_0 + (1-\alpha)A_1}{\cos(\theta_{\rm min}) - \cos(\theta_{\rm max})},
\end{equation}
where
\begin{equation}
\begin{aligned}
A_0 &= \frac{1}{3} \left[\cos^3(\theta_{\rm min}) - \cos^3(\theta_{\rm max})\right], \\
A_1 &= \frac{1}{24}\left[\cos^3(\theta_{\rm max}) - 9\cos(\theta_{\rm max}) - \cos^3(\theta_{\rm min}) + 9\cos(\theta_{\rm min})\right].
\end{aligned}
\end{equation}

\subsubsection{Fitting procedure}
\label{subsect: method2, fitting}

The model was fitted simultaneously to $N_{\rm ord}$ radial and dipole orders over $N_{\rm e}$ epochs, with the same two latitudes for all frequencies resulting in ($2N_{\rm ord} + 2 + N_{\rm e}$) free parameters. Parameters were estimated using a sampling approach where the posterior probability distributions for parameters $\bar{\theta}$, are mapped assuming prior information, $J$, and given data, $D$, using Bayes' theorem:
\begin{equation}
p(\bar{\theta}|D,J) \propto p(D|\bar{\theta},J) \: p(\bar{\theta}|J).
\end{equation}
Here $p(\bar{\theta}|J)$ is the prior, and $p(D|\bar{\theta},J)$ is the likelihood given by
\begin{equation}
p(D|\bar{\theta},J) = \prod\limits_{nl} \left[ \left(2\pi \sigma_{\text{obs},nl}^2\right)^{-\frac{1}{2}} \, \exp \left(\frac{-(\nu_{\text{pred},nl} - \nu_{\text{obs},nl})^2}{2\sigma_{\text{obs},nl}^2} \right) \right],
\label{eqn: likelihood}
\end{equation}
where $\nu_{\text{obs},nl}$ and $\sigma_{\text{obs},nl}$ are the observed frequency and associated uncertainty for a particular mode of order, $n$, and angular degree, $l$, and $\nu_{\text{pred},nl}$ is the modelled frequency of that mode.

Uniform priors were used, with $-5\% \leq\nu_{nl,0}\leq +5\%$, $0\leq\beta_{j}\leq10$ and $0\leq\lambda_{\rm min,max}\leq\frac{\pi}{2}$. The location parameters were constrained so that $\lambda_{\rm min}<\lambda_{\rm max}$, and so that consecutive radial field-free frequencies were separated by the average spacing $\Delta\nu$ between consecutive overtones of the same degree, $l$, with a small tolerance of $\pm 2\,\rm \mu Hz$. Walkers were first made to explore the entire prior space with uniform likelihood so that their starting positions were well spread before fitting our model to the data using the likelihood in Equation \ref{eqn: likelihood}. 

Both the $\beta_j$ and the field-free frequency parameters control the absolute amount of frequency shift, whereas the relative shifts of radial and dipole modes depend on the latitudinal distribution of activity. Since we seek to constrain our determination of the active band latitudes, the actual values of $\beta_j$ are uninformative and we require just the changes in $\beta_j$ over time. The $\beta_j$ parameter corresponding to the first epoch could therefore be constrained using a Gaussian prior, $\mathcal{N}(2.,0.01)$, to speed up convergence. Chosen different values for the first $\beta_j$ made no difference to the resulting latitude predictions. 

\begin{figure}
	\centering
	\includegraphics[width=0.48\textwidth]{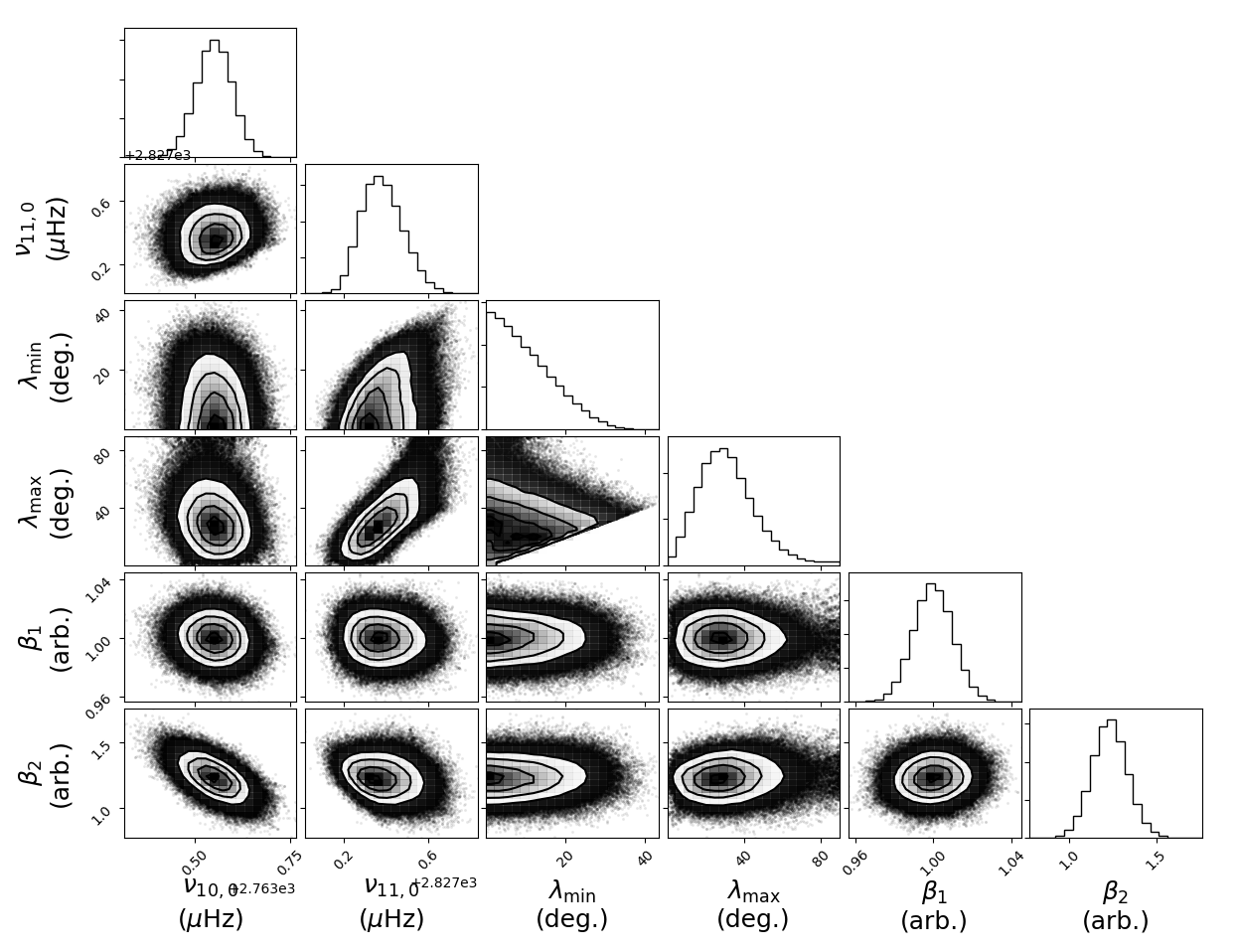}
	\caption{A corner plot showing the joint and marginalised posterior probability distributions for a subset of model parameters fitted to the rising section of the BiSON dataset.}
	\label{fig: corner}
\end{figure}

The fit was made using an affine invariant ensemble sampling algorithm \citep[see][]{2010CAMCS...5...65G} implemented by \texttt{emcee} in Python \citep{2013PASP..125..306F}. Convergence of the chains was checked and the autocorrelation time was calculated to be considerably smaller ($\sim$500 times smaller) than our effective number of samples meaning that the chains were independent of their starting position. Figure \ref{fig: corner} shows an example corner plot \citep{2016JOSS} after convergence with both joint and marginalised posterior distributions for a subset of parameters fitted to the rising phase of the BiSON dataset.
As discussed above, both the field-free frequencies and $\beta_j$ parameters were uninformative and were therefore treated as nuisance parameters. 

\section{Results}
\label{sect: results}

\subsection{The Sun}
\label{subsect: Sun results}

We applied both methods to the BiSON and VIRGO data. Best-fitting latitudes were obtained using the statistical modes of the marginalised posterior distributions, since they were found to be the best summary statistics based on tests with artificial data, i.e., by investigating the response of our top hat model to more realistic-shaped activity distributions with softer boundaries (see below). Reported  uncertainties on the inferred latitudes correspond to the 68 percent credible intervals either side of the mode. 

\begin{figure*}
	\centering
	\begin{subfigure}[]{0.48\textwidth}
		\includegraphics[width=\textwidth]{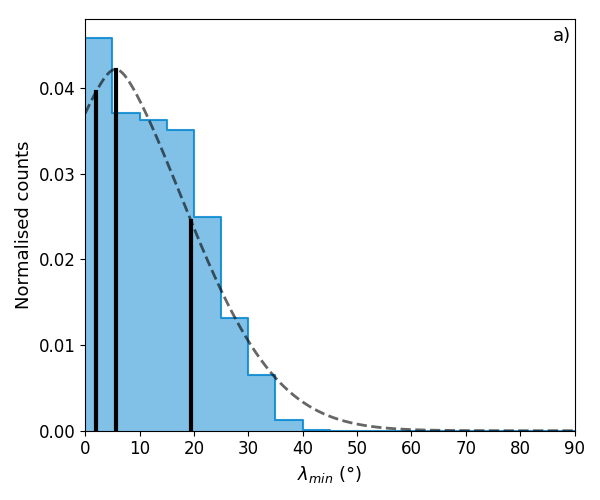}
	\end{subfigure}
	\begin{subfigure}[]{0.48\textwidth}
		\includegraphics[width=\textwidth]{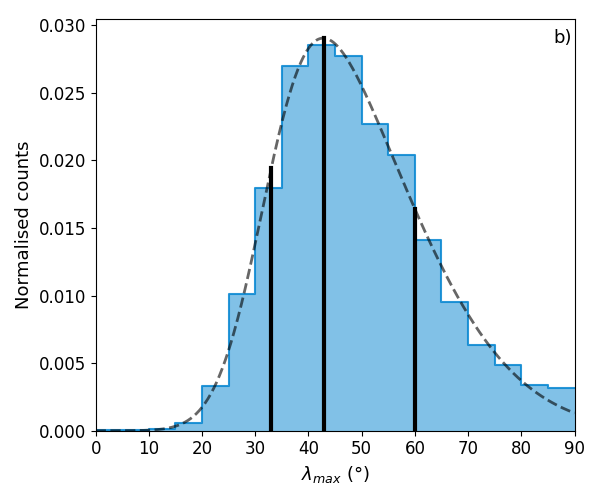}
	\end{subfigure}
	\caption{Posterior distributions of (a) minimum- and (b) maximum-latitudes for the entire 108-day BiSON dataset from method 1. The curves are fitted skewed Gaussians and the vertical lines indicate the mode and 68\,\% confidence intervals.}
	\label{fig: method 1 bison latitudes}
\end{figure*}

\begin{figure*}
	\centering
	\begin{subfigure}[]{0.48\textwidth}
		\includegraphics[width=\textwidth]{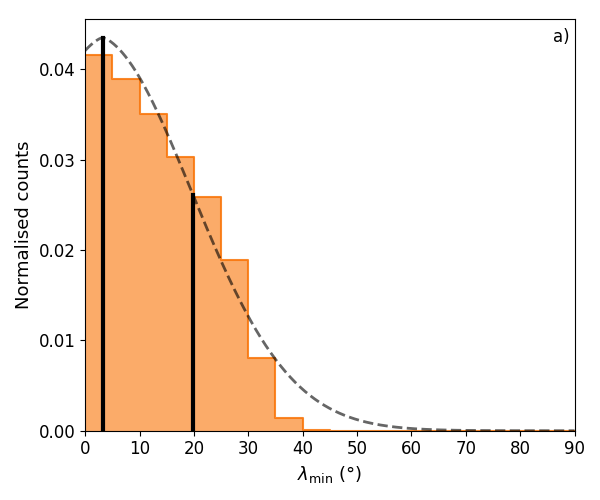}
	\end{subfigure}
	\begin{subfigure}[]{0.48\textwidth}
		\includegraphics[width=\textwidth]{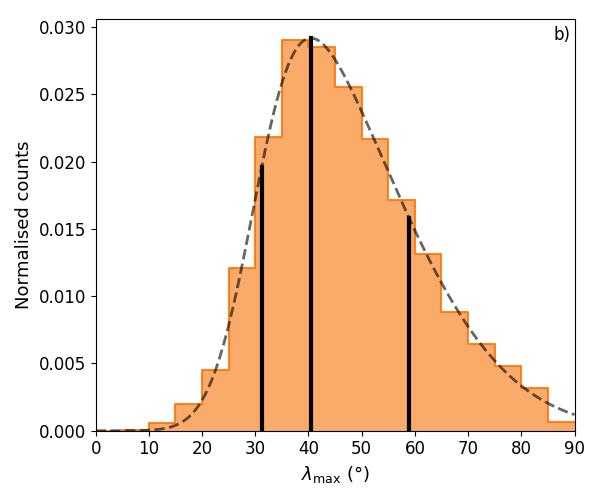}
	\end{subfigure}
	\caption{Posterior distributions of (a) minimum- and (b) maximum-latitudes for the entire 108-day BiSON data from method 2. The curves are fitted skewed Gaussians and the vertical lines indicate the mode and 68\,\% confidence intervals.}
	\label{fig: method 2 bison latitudes}
\end{figure*}

The first method gave estimated minimum latitudes for individual sections of the solar cycle (either rising or falling) of $\sim$10\degree and maximum values of $\sim$40\degree. The latitudes for the BiSON data set covering the entire solar cycle were estimated to be $\lambda_{\rm min} = (5.8^{+13.7}_{-3.8})$\,degrees and $\lambda_{\rm max} = (42.9^{+17.1}_{-9.9})$\,degrees using Method\,1; and $\lambda_{\rm min} = (3.3^{+16.6}_{-16.6})$\,degrees and $\lambda_{\rm max} = (40.6^{+18.3}_{-9.4})$\,degrees using Method\,2. The resulting posterior distributions from the entire observation using each method are shown in Figures \ref{fig: method 1 bison latitudes} and \ref{fig: method 2 bison latitudes}. We obtained similar results using the VIRGO data and other rising and falling phases from BiSON data in addition to those shown in Figure \ref{fig: whole bison cycle}.

In order to measure significant frequency shifts with respect to their uncertainties it was necessary to use frequency differences separated over long periods of time (whole rising or falling sections of the activity cycle). Consequently, the resulting latitudes are an integration over that entire duration. For this reason we cannot expect to detect any migration of activity band latitude during the solar cycle. To be able to detect cyclic variations, frequency shifts would need to be measured over shorter stretches of time which would require either smaller frequency uncertainties and therefore better signal-to-noise levels in the individual power spectra, or larger frequency shifts.

Our results, whilst reassuringly consistent, are notable for the fact that they appear to straddle a wider band in latitude than is usually associated with sunspots. Histograms of spot latitudes shown over recent solar cycles do actually extend beyond 40\,degrees \citep[e.g.][]{2008A&A...483..623S,2017ApJ...851...70M}. However, it is also important to note that our method responds to the range in latitude over which there is significant magnetic flux over and above the weak basal ephemeral flux levels that are present throughout the cycle, as we shall now go on to explain. \citet{2007ApJ...659.1749C} studied the correlation of frequency shifts to proxies of activity that have a different balance of strong- and weaker-field flux, and found that the results implied the frequencies were responding to a mix of strong and weaker-component flux. In addition, \citet{2016MNRAS.461..224S} found that only around 30\% of frequency shifts seem to be directly associated to sunspots, the remaining being likely associated to more global variations, possibly resulting from changes in the overall magnetic field. This supports the idea that the magnetic influence on the frequencies needs not to be restricted to the latitudes where spots are observed.

\begin{figure}
	\centering
	\includegraphics[width=0.48\textwidth]{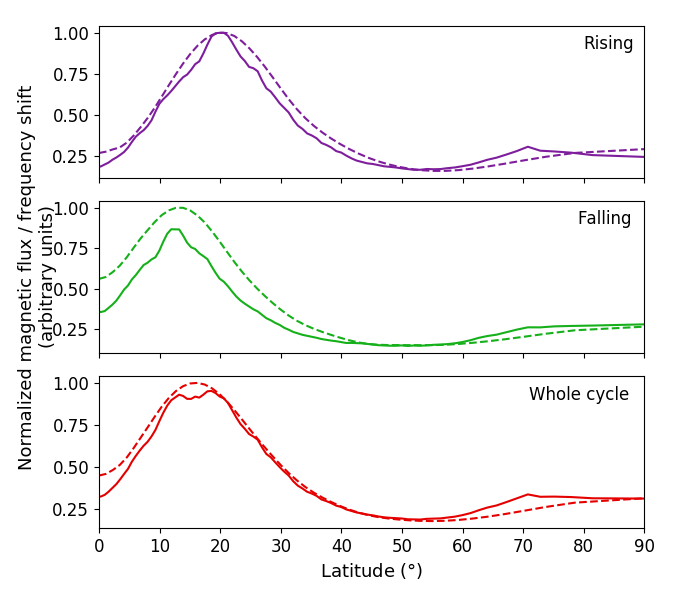}
	\caption{Distribution of mean Carrington rotation averages of magnetic field strength from NSO (solid lines) and frequency shifts from GONG (dashed lines), which have been integrated over rising and falling sections of the solar cycle as defined earlier in the text. }
	\label{fig: gong mag fld}
\end{figure}

Figure \ref{fig: gong mag fld} shows unsigned solar magnetic field strength data from synoptic charts generated by the National Solar Observatory \citep{2017MNRAS.464.4777H}. The solid lines in the top two panels show normalised sums, made over the same two epochs as in our study, of the magnetic flux at different latitudes. These epochs cover the rising and falling phases of cycle 23. The bottom panel shows the sum over both epochs. It is clear from these plots that significant flux extends beyond a latitude of 40\,degrees. As expected, the distributions peak at higher latitudes during the rising phases, since the latitudes at which spots are found moves towards the equator as cycles progress.

Also shown on each plot (dotted lines) is the equivalent latitudinal dependence of frequency shifts shown by medium-$l$ solar p-modes during each epoch \citep{2002ApJ...580.1172H}. These frequencies come from highly spatially resolved observations of the Sun made by the GONG network and, as noted previously, from them it is possible to make inferences on the latitudinal response of the modes to levels of detail that are not possible with the low-$l$ modes only.  As expected, the distributions for the mode frequency shifts follow very closely those of the magnetic flux. We therefore used the flux distributions to generate realistic artificial p-mode data to test our latitudinal inference using the low-$l$ frequencies only. Artificial frequencies were made to mimic six orders of BiSON data straddling the centre of the oscillations spectrum, with $i=90 \pm 4$\,degrees and shifts for different mode components generated using the latitudinal flux distribution integrated over cycle 23. The modelled frequencies covered $N_{\rm e}=12$ and had frequency uncertainties of $\sigma\nu_{\rm obs}=0.15\, \rm \mu Hz$. Frequencies were scattered using a normal distribution with standard deviation equal to the frequency uncertainty. Our top-hat model was then fitted using the previously described methods. 

\begin{figure}
	\centering
	\includegraphics[width=0.48\textwidth]{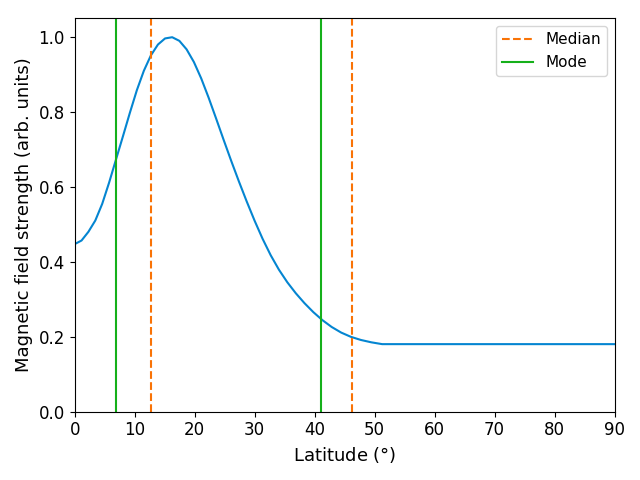}
	\caption{In blue shows the magnetic field distribution used to generate artificial frequency data based on GONG measurements of frequency shifts of the Sun integrated over two solar cycles. Dashed red and solid green lines show the median and modal values, respectively, of $\lambda_{\rm min}$ and $\lambda_{\rm max}$ from fitting our top hat-model.}
	\label{fig: newB bison}
\end{figure}

Figure \ref{fig: newB bison} shows the response of our model to the artificial solar data, showing the median and modal best-fitting latitudes. It is clear that the boundaries of our top-hat model extend to the tails of the latitudinal distribution where the magnetic flux has fallen to ephemeral levels. These tests also show that the modal values better represent the distributions since this test statistic encompasses a much larger proportion of the significant flux.

We verified that our results are robust to using different initial parameter guesses. We also tested artificial data with underlying combinations of $\lambda_{\rm min}$ and $\lambda_{\rm max}$ which covered a range of angles and widths for the active latitude bands, and found that we could recover the expected latitudes given data of the precision available on the Sun and HD\,173701. That the method does not recover the correct \emph{absolute} values of the field-free frequencies and $\beta_j$ coefficients is not a cause for concern. These parameters are highly anti-correlated and we do recover the correct \emph{change} in the $\beta_j$ with time, which allows for robust inference on the latitudes.

\subsection{HD\,173701} 
\label{subsect: doris results}

\begin{figure*}
	\centering
	\begin{subfigure}[]{0.48\textwidth}
		\includegraphics[width=\textwidth]{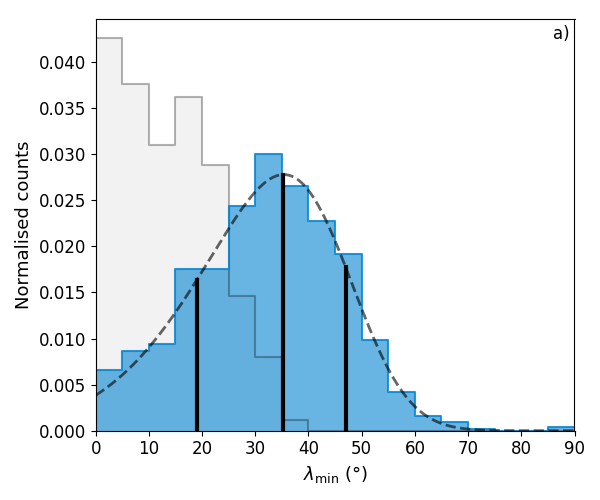}
		\label{fig: method 1 lmin}
	\end{subfigure}
	\begin{subfigure}[]{0.48\textwidth}
		\includegraphics[width=\textwidth]{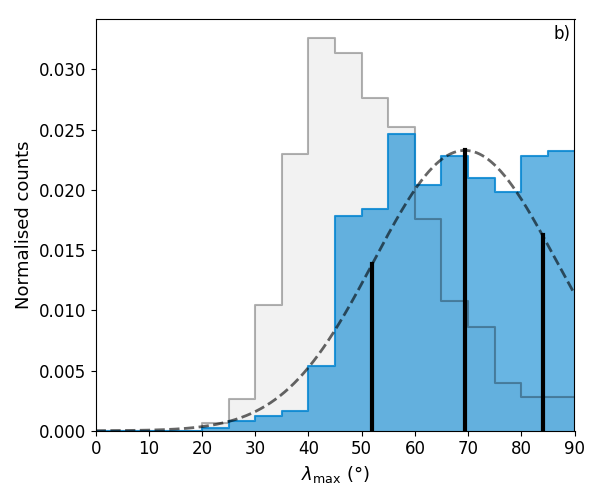}
		\label{fig: method 1 lmax}
	\end{subfigure}
	\caption{Posterior distributions of (a) minimum- and (b) maximum-latitudes for HD\,173701 from method 1. In faded grey is the distribution for BiSON analysis over an equivalent duration of the solar cycle (4.4\,years) for comparison. The curves are fitted skewed Gaussians and the vertical lines indicate the mode and 68\,\% confidence intervals.}
	\label{fig: method 1 doris latitudes}
\end{figure*}

\begin{figure*}
	\centering
	\begin{subfigure}[]{0.48\textwidth}
		\includegraphics[width=\textwidth]{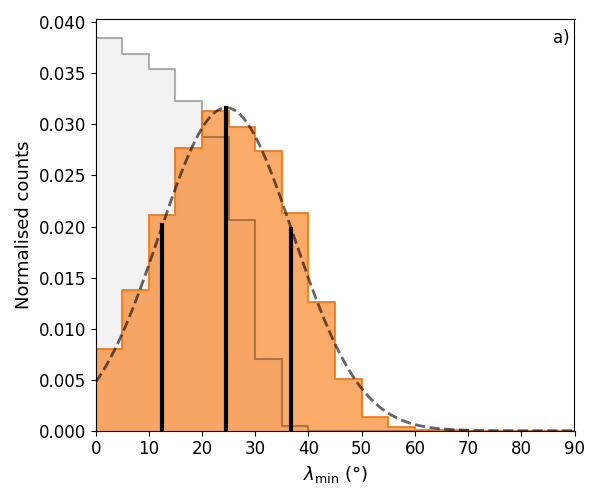}
	\end{subfigure}
	\begin{subfigure}[]{0.48\textwidth}
		\includegraphics[width=\textwidth]{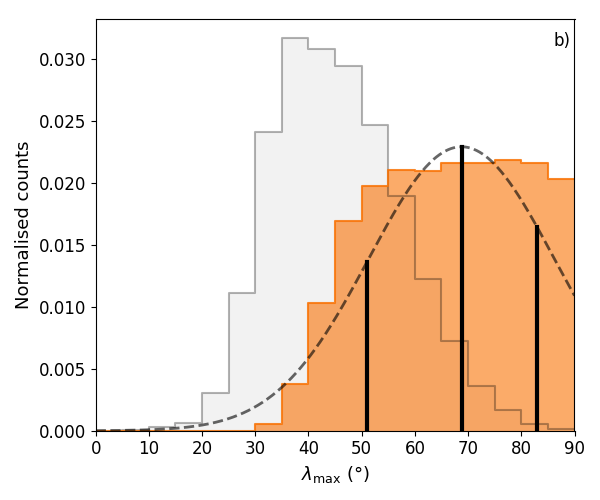}
	\end{subfigure}
	\caption{Posterior distributions of (a) minimum- and (b) maximum-latitudes for HD\,173701 from method 2. In faded grey is the distribution for BiSON analysis over an equivalent duration of the solar cycle (4.4\,years) for comparison. The curves are fitted skewed Gaussians and the vertical lines indicate the mode and 68\,\% confidence intervals.}
	\label{fig: method 2 doris latitudes}
\end{figure*}

Both methods were applied to the 90-day \textit{Kepler} datasets on HD\,173701. The posterior distributions on the latitudes given by the first and second methods are shown in Figures \ref{fig: method 1 doris latitudes} and \ref{fig: method 2 doris latitudes} respectively. The inferred latitudes for HD\,173701 are $\lambda_{\rm min} = (34.5^{+11.5}_{-16.5})$\,degrees and $\lambda_{\rm max} = (69.6^{+13.9}_{-17.1})$\,degrees from Method\,1; and $\lambda_{\rm min} = (24.9^{+12.0}_{-12.2})$\,degrees and $\lambda_{\rm max} = (68.8^{+14.2}_{-17.8})$\,degrees from Method\,2. As with the solar results, the two separate analyses produced consistent conclusions for HD\,173701.

In order to make comparisons with the Sun, a section of BiSON data was chosen to span the same proportion of the star's activity cycle. This corresponded to 4.4 years of BiSON data, since the cycle of HD\,173701 is 7.4 years, of which we have observed 3.0 years. We know that the amplitudes of the frequency shifts in HD173701 are larger than those in the Sun and so it should be easier for the model to predict the latitudes of activity given that the power-spectra signal-to-noise ratios are similar. 

It is clear from comparison between the latitude posterior distributions using the equivalent duration from BiSON (shown in grey) that the spread of activity on HD\,173701 is quite different to that on the Sun and is more widely distributed, i.e. it extends to much higher latitudes. From the application of a Kolmogorov-Smirnov test \citep{KolmogorovA1933,SmirnovN1939}, we may conclude that the probability is vanishingly small that the $\lambda_{\rm min,max}$ posteriors for the Sun and HD\,173701 are drawn from the same underlying distribution.

We again performed tests with artificial data, this time with simulated frequencies made from an underlying model of the magnetic flux similar to that shown in Figure \ref{fig: newB bison} but peaked at  higher degrees (around 55\,degrees). We again found that our results had sensitivity extending to the tails of the modelled magnetic field distribution.

\section{Conclusions}
\label{sect: conclusions}

We have presented a new approach to constraining active latitudes on solar-type stars, which uses stellar-cycle-induced frequency shifts of detectable solar-like oscillations. These oscillations may be described by spherical harmonic functions of angular degree $l$ and azimuthal order $m$. When near-surface activity is distributed in a non-homogeneous manner, modes of different ($l$,$m$) will have their frequencies shifted by different amounts. By adopting a simple model for the spatial distribution of the near-surface activity, we show that data on the frequency shifts may be used to infer minimum and maximum latitudes for the active bands, $\lambda_{\rm min}$ and $\lambda_{\rm max}$, respectively.

We presented two complementary methods, which we tested on Sun-as-a-star helioseismic data and artificial data. From BiSON data covering solar cycle 23, we found $\lambda_{\rm min} = (5.8^{+13.7}_{-3.8})$\,degrees and $\lambda_{\rm max} = (42.9^{+17.1}_{-9.9})$\,degrees using our first method; and $\lambda_{\rm min} = (3.3^{+16.6}_{-16.6})$\,degrees and $\lambda_{\rm max} = (40.6^{+18.3}_{-9.4})$\,degrees using our second method. We obtained similar results using data from the VIRGO/SPM instrument on board the ESA/NASA SoHO satellite. At first glance, these solar results appear to straddle a wider band in latitude than is usually associated with sunspots. However, histograms of spot latitudes do extend beyond 40\,degrees over recent cycles. Moreover, we also showed that our method responds to the range in latitude over which there is significant magnetic flux over and above the weak basal ephemeral flux levels that are present throughout the cycle.

We then applied our methods to just over 3\, years of \emph{Kepler} data on the solar analogue HD\,173701. This star shows a 7.4-year activity cycle, and stronger frequency shifts than the Sun. We found $\lambda_{\rm min} = (34.5^{+11.5}_{-16.5})$\,degrees and $\lambda_{\rm max} = (69.6^{+13.9}_{-17.1})$\,degrees from our first method; and $\lambda_{\rm min} = (24.9^{+12.0}_{-12.2})$\,degrees and $\lambda_{\rm max} = (68.8^{+14.2}_{-17.8})$\,degrees from our second method. Comparing the peaks of the posterior distributions it is clear that the active bands on HD\,173701 straddle a much wider range in latitude than do the bands on the Sun. From our inferred posteriors, we find that there is a vanishingly small chance that the results on HD\,173701 and the Sun describe the same underlying distribution.

We may compare our findings with those from spot modulation signatures found in \emph{Kepler} data on this star. \citet{2018ApJ...852...46K} inferred spot latitudes ranging from a few up to 40\,degrees, using the relative strengths of different rotational harmonics of the spot modulation signal. \citet{2018A&A...619L...9B} combined the spot modulation signal with asteroseismic constraints on latitudinal differential rotation given by the frequency splittings of non-radial p modes, and uncovered similar latitudes to \citet{2018ApJ...852...46K}, but with results plotted over time. 

That these two studies find similar results is perhaps not surprising, given that both use the same spot modulation signal. Such results are of course sensitive to the latitudes of discrete, strong-field structures, i.e., spots. A possible explanation for why our inference on the activity extends over a much wider latitude range is that our results are recording the presence of significant amounts of more widely distributed, weaker-component flux, lying outside spots.

\section*{Acknowledgements}
We thank Dr R. Komm (National Solar Observatory) for the tabulation of the magnetic field-strength data. A.E.L.T., W.J.C., G.R.D., R.H. and Y.E. would acknowledge the support of the Science and Technology Facilities Council (STFC). Funding for the Stellar Astrophysics Centre is provided by The Danish National Research Foundation (Grant agreement no.:DNRF106). A.R.G.S. acknowledges the support from National Aeronautics and Space Administration under Grant NNX17AF27G. A.M. acknowledges the European Research Council (ERC) under the European Union's Horizon 2020 research and innovation programme (project ASTEROCHRONOMETRY, grant agreement no. 772293). T.L.C. acknowledges support from the European Union's Horizon 2020 research and innovation programme under the Marie Sk\l{}odowska-Curie grant agreement No.792848. M.S.C. is supported by Funda\c{c}\~{a}o para a Ci\^{e}ncia ea Tecnologia (FCT) through national funds and by FEDER through COMPETE2020 in connection to these grants: UID/FIS/04434/2013 \& POCI-01-0145-FEDER-007672, PTDC/FIS-AST/30389/2017 \& 
POCI-01-0145-FEDER-030389.

This work also made use of \texttt{emcee} \citep{2013PASP..125..306F} as well as the open-source Python packages Numpy \citep{2015NUMPY}, Scipy \citep{SCIPY}, Pandas \citep{2010mckinney-proc-scipy}, Matplotlib \citep{2007MATPLOTLIB}.




\bibliographystyle{mnras}
\bibliography{main} 



\appendix
\section{Frequency dependence on inclination angle}
\label{adx: alpha_i_relation}
As discussed in Section \ref{sect: general theory}, we must in practice work with non-radial mode frequencies which represent some suitably weighted average of the frequencies of the constituent $m$ components.

The relative visibilities of the individual azimuthal components of a mode depend on the angle of inclination shown by the star. In the case where it is difficult to place tight constraints on the frequencies of the individual $m$ components -- e.g., when shorter datasets are needed to follow stellar-cycle changes in the frequencies over time -- it is necessary to understand the relative contribution of the constituent $m$ to the central frequency of the multiplet. Here, we place constraints on those contributions for $l=1$ modes.

The individual components of a non-radial mode are split by rotation. Moreover, because the activity-induced shift at each ($l$,$|m|$) is likely to differ in size, the result is a multiplet in which the constituent frequencies are arranged asymmetrically. At $l=1$, the component frequencies are related by
\begin{equation*}
\begin{aligned}
\nu_{10} &= \nu_0, \\
\nu_{11} &= \nu_0 + \nu_{\rm s},\\
\nu_{1-1} &= \nu_0 - \nu_{\rm s} +2a.\\
\end{aligned}
\end{equation*}
where $\nu_0$ is the frequency all components would have in the absence of rotation or magnetic fields, $\nu_{\rm s}$ is the rotational splitting, and $a$ is the frequency asymmetry.

\begin{figure}
	\centering
	\includegraphics[width=0.5\textwidth]{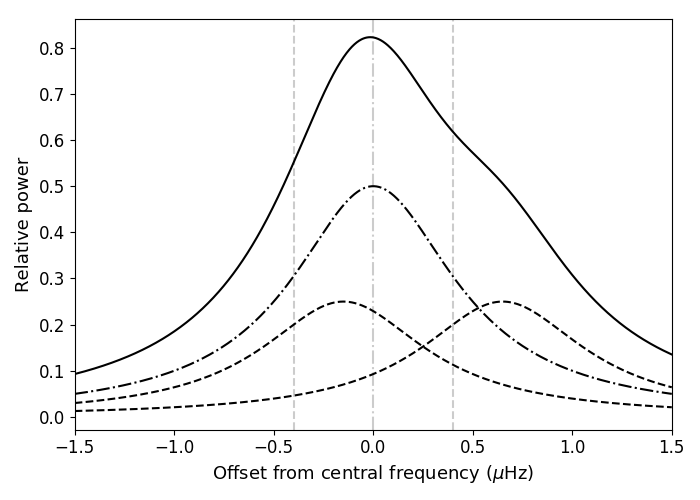}
	\caption{Multiplet peak profile of an $l=1$ mode situated at $i=45\degree$. The dot-dashed curve indicates $m=0$, dashed curve $|m|=1$, the solid curve the combined power, and vertical lines with corresponding line style indicate the positions of the individual azimuthal components in the absence of the perturbing magnetic field.}
	\label{fig: model multiplet}
\end{figure}

Figure \ref{fig: model multiplet} shows a representation of the limit profile in frequency for a multiplet viewed at $i=45\degree$, having Lorentzian peak linewidths of $1\mu$Hz, a rotational splitting of $\nu_{\rm s}=0.4\,\rm \mu Hz$ and an asymmetry of $a=0.25\,\rm \mu Hz$. The mode parameters are typical of a prominent mode in the solar oscillations spectrum for high levels of activity; these parameters are also representative of HD\,173701, whilst the angle chosen for the plot is similar to that extracted on HD\,173701. The two $|m|=1$ components, shown by the dashed lines, have been shifted from the positions they would have occupied in the absence of the near-surface magnetic field (as shown by the grey vertical lines). The overall result is a slightly distorted multiplet whose centroid has been shifted.

Artificial data were generated for $l=1$ modes comprising three azimuthal components, plus background, with the combined power spectral density described by
\begin{equation}
M(\nu;\nu_{lm}) = \sum\limits_{m=-l}^l \epsilon_{lm}(i) \, M_L(\nu;\nu_{lm}) + B,
\end{equation}
where the component frequencies are $\nu_{lm}$, $\epsilon_{lm}(i)$ is the visibility,  $B$ the background, and $M_L(\nu_{lm})$ describes a Lorentzian profile with the form 
\begin{equation}
M_L(\nu;\nu_{lm}) = \frac{H}{1 + \frac{4}{\Gamma^2} (\nu - \nu_{lm})^2},
\end{equation}
where $\Gamma$ is the mode linewidth,  and the mode height is $H$.

We used $\Gamma=1\,\mu$Hz and $\nu_s=0.4\,\mu$Hz. The generated data were given a frequency resolution and height-to-background ratio, $H/B$, representative of a \textit{Kepler} 90-day dataset. We made data for different inclination angles, covering the range zero to 90\,degrees in steps of 0.5\,degrees, and different asymmetries $a$ in the range $(-1.0\leq a\leq1.0)\,\mu$Hz at increments of $0.2\mu$Hz. Noise was included by perturbing the limit spectra with negative exponential statistics \citep{1990ApJ...364..699A}.

A symmetric model was fitted to the generated data using Markov chain Monte Carlo (MCMC) techniques \citep[e.g.][]{2010CAMCS...5...65G}. The parameters $H$, $\Gamma$, $\nu_s$, $i$ and $B$ were set as their true values for all fits and the process was run to find $\nu_0$ using 100 simultaneous walkers and 5000 iterations. For a single inclination angle and asymmetry this was repeated 50 times and a total posterior distribution constructed over all fits and iterations. The median gave the fitted frequency, $\nu_{\rm fit}$, and the uncertainties were taken from 68\,\% confidence intervals. We found that small changes to the underlying mode parameters produced almost identical results for the fitted centroid frequencies.

\begin{figure}
	\centering
	\includegraphics[width=0.5\textwidth]{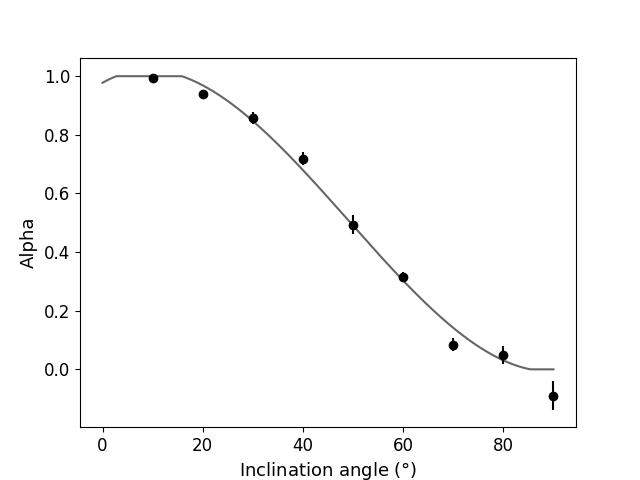}
	\caption{Plot of $\alpha$ versus inclination angle found using methods described in the text. The solid line shows the spline fit used to find $\alpha$ at any angle, $i$.}
\label{fig: alpha vs i}
\end{figure}

By fitting a symmetric model to asymmetric data we tested how much the central frequency of an $l=1$ mode is affected by fitting an inaccurate symmetric model. This depends on the balance of each azimuthal contribution and therefore on the inclination angle. We define a quantity $\alpha$ which describes the relative contributions from $m=0$ and $|m|=1$ components, $\nu_{10}$ and $\nu_{1|1|}$ respectively, to the fitted frequency in the equation
\begin{equation}
\nu_{\rm fit} = \alpha \nu_{10} + (1-\alpha)\nu_{1|1|},
\label{eqn: nufit}
\end{equation}
where 
\begin{equation}
\nu_{1|1|}= \frac{1}{2}(\nu_{11}+\nu_{1-1}) \equiv \nu_0 + a.
\end{equation}
Rearranging Equation \ref{eqn: nufit} gives
\begin{equation}
\alpha = \frac{\nu_{\rm fit} - \nu_{1|1|}}{\nu_0 - \nu_{1|1|}} = \frac{\nu_{\rm fit} - \nu_{1|1|}}{-a}.
\end{equation}
Therefore if we plot $(\nu_{\rm fit} - \nu_{1|1|})$ against $a$, the negative gradient will give $\alpha$. Values of $\alpha$ along with their uncertainties were calculated for each inclination angle. A spline fit to this relation enabled interpolation for any other angle (see Figure \ref{fig: alpha vs i}). 

\bsp	
\label{lastpage}
\end{document}